\documentstyle[11pt,a4]{article}
\renewcommand{\epsilon}{\varepsilon}
\begin{document}
\title{Cosmological two-fluid thermodynamics}

\author{Winfried Zimdahl\footnote{Fakult\"at f\"ur Physik, Universit\"at Konstanz,
PF 5560 M678, 
D-78457 Konstanz, Germany, electronic address: winfried.zimdahl@uni-konstanz.de} 
and Diego Pav\'{o}n\footnote{Departamento de F\'{\i}sica
Universidad Aut\'{o}noma de Barcelona,
08193 Bellaterra (Barcelona), Spain 
electronic address:  
diego@ulises.uab.es}}

\date{\today}
\maketitle
%\thispagestyle{empty}
%\pacs{}

\begin{abstract}
We reveal unifying thermodynamic aspects of so 
different phenomena as the cosmological electron-positron annihilation, the evaporation of primordial black holes with a narrow mass range, and the ``deflationary'' transition from an initial de Sitter phase to a subsequent standard Friedmann-Lema\^{\i}tre-Robertson-Walker 
(FLRW) behavior.   
\end{abstract}

\section{Introduction}
\label{Introduction}
The thermodynamics of two fluids with different temperatures represents a framework which is sufficiently general to apply to entirely different epochs of the cosmological evolution. 
This unifying feature may be used to establish surprising similarities between otherwise quite independent phenomena in the expanding universe. 
In this paper we focus on aspects of the temperature evolution during periods with decay and production of particles to demonstrate the universal power of the thermodynamic description. 
In particular, we show that the same simple law for the cooling rate of a fluid in the expanding universe governs a wide range of phenomena implying the cosmological electron-positron annihilation after neutrino decoupling at about 1 MeV, the evaporation of primordial black holes (PBHs) with a narrow mass range, and the ``deflationary'' \cite{Barrow} transition from an initial de Sitter stage to a subsequent FLRW period, equivalent to a phenomenological vacuum decay model. 
All these processes are characterized by a strong back reaction of decay and production processes on the thermal evolution of the universe. 
It is the possibility of taking into account this back reaction in a rather straightforward but general way, which admits an application to such a variety of different phenomena. 

More specifically, we shall first reproduce the factor 
$\left(11/4 \right)^{1/3}$ by which the temperature of the neutrino background differs from that of the photon background as a consequence of electron-positron annihilation. 
Secondly, we show that the black hole temperature behavior 
$T _{_{\left(BH \right)}}\propto m _{_{\left(BH \right)}}^{-1}$, where 
$m _{_{\left(BH \right)}}$ is the black hole mass, is consistent with  
the general fluid temperature law for a PBH ``fluid'', a configuration in which all its members are assumed to have the same mass $m _{_{\left(BH \right)}}$.   On this basis we discuss thermodynamical aspects of PBH evaporation. 
The third example is the evolution of the radiation temperature in a ``deflationary'' scenario of the early universe which implies an initial increase to a maximum value as a result of the production of relativistic particles out of a decaying vacuum, followed by a decrease which finally approaches the familiar FLRW behavior. 

None of these results is really new. 
The first case is cosmological textbook physics (see, e.g. \cite{Boe}), the second one was investigated in \cite{ZPPBH}, the scenario characterizing the third case is based on \cite{GunzMaNe} (see also \cite{LiMa,Zpreprint}). 
What is new however, is the unifying view which allows the discussion of so different cosmological effects starting from the same set of basic equations. 
To highlight the underlying common thermodynamical features of the mentioned phenomena is the main purpose of this paper.    
The material is organized as follows. 
In section \ref{basic} we recall the basic relations of two-fluid thermodynamics in an expanding universe. These relations are used in section \ref{epa} to discuss the cosmological electron-positron annihilation. 
In section \ref{PBH} the general formalism is applied to a mixture of radiation and a component of PBHs which are assumed to have the same mass. 
It may be shown that under this condition they share essential properties with a pressureless gas. 
Thermodynamic aspects of a smooth transition from a de Sitter stage to a radiation dominated FLRW phase including an intermediate temperature maximum 
are investigated on the same general basis in section \ref{defl}, while the final section \ref{conclusions} is devoted to  concluding remarks.   
Units have been chosen so that 
$c = k_{B} = \hbar = 1$.

\section{Basic thermodynamic relations}
\label{basic}

We assume the cosmic medium to consist of two components which share the same 4-velocity $u ^{i}$. Each of the components has a perfect fluid structure with the energy-momentum tensor 
$T ^{ik}_{_{\left(A \right)}}$, where $A=1,2$, and a corresponding particle flow vector $N ^{i}_{_{\left(A \right)}}$,  
\begin{equation}
T^{ik}_{_{\left(A \right)}} = \rho_{_{\left(A \right)}} u^{i}u^{k}
+ p_{_{\left(A \right)}} h^{ik}{\mbox{ ,}} \; \; \;
N_{_{\left(A \right)}}^{i} = n_{_{\left(A \right)}}u^{i}{\mbox{ , }}
\mbox{\ \ \ }\mbox{\ \ \ }(A = 1, 2)\ .
\label{1}
\end{equation}
Here, $\rho _{_{\left(A \right)}}$ is the energy density of component $A$, measured by a comoving observer, $p _{_{\left(A \right)}}$  is the corresponding equilibrium pressure,  
$h ^{ik}=g ^{ik}+u ^{i}u ^{k}$ is the spatial projection tensor, and 
$n _{_{\left(A \right)}}$ is the number density of species-$A$ particles.  
Neither $T ^{ik}_{_{\left(A \right)}}$ nor 
$N ^{i}_{_{\left(A \right)}}$ are required to be conserved, i.e., interactions and interparticle conversions are admitted: 
\begin{equation}
T ^{ik}_{_{\left(A \right)} ;k}
= - t _{_{\left(A \right)}}^{i} {\mbox{ , }} 
\mbox{\ \ }\mbox{\ \ }\mbox{\ \ }  
N _{_{\left(A \right)} ;i}^{i}
= \dot{n}_{_{\left(A \right)}}
+ 3H n_{_{\left(A \right)}} 
= n _{_{\left(A \right)}}\Gamma _{_{\left(A \right)}}\ .
\label{2}
\end{equation}
The quantity $H$ is the Hubble parameter $H=\dot{a}/a$  with the scale factor $a$ of the Robertson-Walker metric. 
$\Gamma _{_{\left(A \right)}}\equiv  
\dot{N}_{_{\left(A \right)}}/N _{_{\left(A \right)}}$ denotes the rate of change of the number 
$N _{_{\left(A \right)}}\equiv  n _{_{\left(A \right)}}a ^{3}$ of particles in a comoving volume $a ^{3}$.  
The $T ^{ik}_{_{\left(A \right)}}$ and $
N ^{i}_{_{\left(A \right)}}$ add up to the corresponding quantities for the medium as a whole: 
\begin{equation}
T ^{ik} = T ^{ik}_{_{\left(1 \right)}}
+ T ^{ik}_{_{\left(2 \right)}} {\mbox{ , }} 
\mbox{\ \ \ }\mbox{\ \ \ }
N ^{i} = N _{_{\left(1 \right)}} + N _{_{\left(2 \right)}}\ .
\label{3}
\end{equation}
It is well known that in general the energy-momentum tensor $T ^{ik}$ does not take the form of a perfect fluid, but will contain a non-equilibrium pressure $\Pi $  \cite{UI,ZMN96,ZMN97}. 
Different from the $T ^{ik}_{_{\left(A \right)}}$, the overall energy-momentum tensor has to be conserved, which establishes a relation between 
$t ^{i}_{_{\left(1 \right)}}$ and $t ^{i}_{_{\left(2 \right)}}$: 
\begin{equation}
T^{ik} = \rho u^{i}u^{k} 
+ \left(p + \Pi\right) h^{ik} \ ,
\mbox{\ \ } 
\mbox{\ \ }
T^{ik}_{\ ;k} = 0 
\quad\Rightarrow\quad 
t ^{i}_{_{\left(1 \right)}} = - t ^{i}_{_{\left(2 \right)}}\ .
\label{4}
\end{equation}
We do not require, however, conservation of the total particle number \cite{Prig,Calv}, i.e., 
\begin{equation}
N^{a}_{;a}  =  
\dot{n} + 3H  n =  n \Gamma  
\equiv  n _{_{\left(1 \right)}}\Gamma _{_{\left(1 \right)}} 
+ n _{_{\left(2 \right)}}\Gamma _{_{\left(2 \right)}}
\ , 
\label{5}
\end{equation}
where $n \equiv  n _{_{\left(1 \right)}}+ n _{_{\left(2 \right)}}$ is the overall particle number density. 
Each component is governed by its own Gibbs equation which provides us with an expression for the time behaviour of the entropy per particle 
$s _{_{\left(A \right)}}$, 
\begin{equation}
T _{_{\left(A \right)}} \mbox{d} s _{_{\left(A \right)}}
= \mbox{d} \frac{\rho _{_{\left(A \right)}}}
{n _{_{\left(A \right)}}}
+ p _{_{\left(A \right)}}
\mbox{d} \frac{1}{n _{_{\left(A \right)}}} {\mbox{ , }}
\quad\Rightarrow\quad
n _{_{\left(A \right)}} T _{_{\left(A \right)}}
\dot{s}_{_{\left(A \right)}} = u _{a} t _{_{\left(A \right)}}^{a}
- \left(\rho _{_{\left(A \right)}}
+ p _{_{\left(A \right)}}\right) \Gamma _{_{\left(A \right)}} {\mbox{ .}}
\label{6}
\end{equation}
In general, the temperatures $T _{_{\left(A \right)}}$ of both components are different. 
With the help of the equations of state 
\begin{equation}
p_{_{\left(A \right)}} = p_{_{\left(A \right)}}
\left(n_{_{\left(A \right)}}, T_{_{\left(A \right)}}\right) \ ,
\mbox{\ \ }
\rho_{_{\left(A \right)}} = \rho_{_{\left(A \right)}}
\left(n_{_{\left(A \right)}},
T_{_{\left(A \right)}}\right){\mbox{ , }}
\label{7}
\end{equation}
one obtains the evolution law for the temperatures $T _{_{\left(A \right)}}$. 
Namely, differentiating $\rho _{_{\left(A \right)}}$ in (\ref{7}) 
along the fluid flow lines and applying the balances (\ref{2}) we find 
\cite{ZMN97,Calv,LiGer}
\begin{equation}
\frac{\dot{T}_{_{\left(A \right)}}}{T_{_{\left(A \right)}}} = - 
3H\left(1 - \frac{\Gamma _{_{\left(A \right)}}}{3H} \right)
\frac{\partial p_{_{\left(A \right)}}}
{\partial \rho_{_{\left(A \right)}}}
+ \frac{n _{_{\left(A \right)}}
\dot{s}_{_{\left(A \right)}}}
{\partial \rho _{_{\left(A \right)}}/ \partial T _{_{\left(A \right)}}}
{\mbox{ ,}}
\label{8}
\end{equation}
where 
\[
\frac{\partial{p _{_{\left(A \right)}}}}
{\partial{\rho _{_{\left(A \right)}}}} \equiv
\frac{\left(\partial p_{_{\left(A \right)}}
/\partial T_{_{\left(A \right)}} \right)_{n _{_{\left(A \right)}}}}
{\left(\partial \rho_{_{\left(A \right)}}/
\partial T_{_{\left(A \right)}} \right)_{n _{_{\left(A \right)}}}} \ ,
\mbox{\ \ }
\frac{\partial \rho_{_{\left(A \right)}}}{\partial T_{_{\left(A  
\right)}}}
\equiv   \left(\frac{\partial \rho_{_{\left(A \right)}}}{\partial  
T_{_{\left(A \right)}}} \right)_{n _{_{\left(A \right)}}}\ .
\]
The general temperature law (\ref{8}) provides the unifying basis for the discussions of the following sections. It will play a central role in our investigations of both the electron-positron annihilation and the PBH evaporation and a specific inflationary scenario. 

An important special case which we frequently will refer to is characterized by the condition $\dot{s}_{_{\left(A \right)}} = 0$, which means constant entropy per particle \cite{Prig,Calv}. 
This condition simplifies the temperature law,  
\begin{equation}
\dot{s}_{_{\left(A \right)}} = 0 
\quad\Rightarrow\quad
\frac{\dot{T}_{_{\left(A \right)}}}{T_{_{\left(A \right)}}}  = -  3H
\left(1 - \frac{\Gamma _{_{\left(A \right)}}}{3H}
 \right)
\frac{\partial{p _{_{\left(A \right)}}}}
{\partial{\rho _{_{\left(A \right)}}}}
{\mbox{ .}}  
\label{9}
\end{equation}
Moreover, according to (\ref{6}) it establishes the link 
$u _{a} t _{_{\left(A \right)}}^{a}
= \left(\rho _{_{\left(A \right)}}
+ p _{_{\left(A \right)}}\right) \Gamma _{_{\left(A \right)}}$ 
between the source terms 
$\Gamma _{_{\left(A \right)}}$ and $t ^{i}_{_{\left(A \right)}}$ which together with the last relation of 
(\ref{4}) provides us with a relation between the rates 
$\Gamma _{_{\left(1 \right)}}$ and $\Gamma _{_{\left(2 \right)}}$: 
\begin{equation}
n _{_{\left(2 \right)}}\Gamma _{_{2}} 
= - \frac{h _{_{\left(1 \right)}}}{h _{_{\left(2 \right)}}}
n _{_{\left(1 \right)}}\Gamma _{_{\left(1 \right)}}
\quad\Rightarrow\quad
\dot{N}_{_{\left(2 \right)}} = 
-\frac{h_{_{\left(1 \right)}}}{h_{_{\left(2 \right)}}}
\dot{N}_{_{\left(1 \right)}}{\mbox{ }} 
\mbox{\ \ \ }\mbox{\ \ \ }\mbox{\ \ \ }\mbox{\ \ \ }\mbox{\ \ \ }
\left(\dot{s}_{_{\left(A \right)}} = 0 \right)\ ,
\label{10}
\end{equation}
where  $h _{_{\left(A \right)}} \equiv 
\left(\rho _{_{\left(A \right)}} + p _{_{\left(A \right)}}
\right)/ n _{_{\left(A \right)}}$ are the enthalpies per particle.

\section{Electron-positron annihilation}
\label{epa}

Let us consider the cosmological period of 
electron-positron annihilation and the 
corresponding creation of photons, 
shortly after neutrino decoupling  
(see, e.g. \cite{Boe}, \S 3.1.2). 
Annihilation becomes predominant as soon as the radiation temperature drops below the mass of the electron. 
Before the electron-positron annihilation 
there are two bosonic degrees
of freedom (photons), four fermionic degrees 
of freedom due to the
electrons and positrons, and 12 fermionic degrees of 
freedom due to the
different neutrino species.
The fermionic energy density is 
\begin{equation}
\rho_{F} = \frac{7}{6}\frac{\pi^{4}}{30\zeta\left(3\right)}
n_{F}T _{F}{\mbox{ , }}
\ \ 
\mbox{\ \ \ }
n_{F} = \frac{3}{4}\frac{\zeta\left(3\right)}
{\pi^{2}}g_{F}T _{F}^{3}
{\mbox{ , }}
\label{12}
\end{equation}
where $\zeta \left(x\right)$ is Riemann's Zeta-function, 
and
the bosonic one
\begin{equation}
\rho_{B} = \frac{\pi^{4}}{30\zeta\left(3\right)}
n_{B}T _{B}{\mbox{ , }} \ \ 
\mbox{\ \ \ }
n_{B} = \frac{\zeta\left(3\right)}{\pi^{2}}g_{B}T _{B}^{3}
{\mbox{ .}}
\label{13}
\end{equation}
The factors $g_{F}$ and $g_{B}$ are the numbers of fermionic 
and bosonic degrees of
freedom, respectively.

After the electron-positron annihilation we are left 
with the two
bosonic degrees of freedom (photons) and the 
12 neutrino degrees of
freedom, i.e., four fermionic degrees of freedom 
have disappeared. 
The neutrino degrees of freedom are not affected at 
all by this process.
The neutrino temperature behaves as 
$T _{_{\left(\nu  \right)}}=T _{_{\left(\nu  \right)}}\left(t _{0} \right)
a \left(t _{0}\right)/a \left(t \right)$, where the initial time 
$t _{0}$  is assumed to be a time before the beginning of the annihilation process, i.e., electrons, positrons and photons are still at equilibrium at 
$t _{0}$ with 
$T _{_{\left(\nu  \right)}}\left(t _{0} \right)
=T _{_{\left(e ^{\pm }  \right)}}\left(t _{0} \right)
=T _{_{\left(\gamma  \right)}}\left(t _{0} \right)
=T _{_{\left(0 \right)}}$.  
Let us consider the subsystem of four 
fermionic and two
bosonic degrees of freedom \cite{ZPepa}. 
The four fermionic degrees of freedom due to the 
electrons and positrons
are identified with fluid 1 of our general analysis, 
i.e., $\left(1 \right)\rightarrow \left(e ^{\pm } \right)$ 
while the two
bosonic degrees of freedom due to the photons are fluid 2, i.e., 
$\left(2 \right)\rightarrow \left(\gamma  \right)$.  
With
these specifications the number densities become 
\begin{equation}
n_{_{\left(e ^{\pm } \right)}} = 4 \frac{3}{4}\frac{\zeta\left(3\right)}{\pi^{2}}T_{_{\left(e ^{\pm } \right)}}^{3}
{\mbox{ , }}
\mbox{\ \ \ }
n_{_{\left(\gamma  \right)}} 
= 2 \frac{\zeta\left(3\right)}{\pi^{2}}T_{_{\left(\gamma  \right)}}^{3}
{\mbox{ .}}
\label{14}
\end{equation}
The corresponding enthalpies per particle are 
\begin{equation}
h _{_{\left(e ^{\pm } \right)}} 
= \frac{14}{9} \frac{\pi ^{4}}{30 \zeta \left(3\right)}
 T _{_{\left(e ^{\pm } \right)}}
{\mbox{ , }}
\ \ 
\mbox{\ \ \ }
h _{_{\left(\gamma  \right)}} =  \frac{4}{3}
\frac{\pi ^{4}}{30 \zeta \left(3\right)}
T _{_{\left(\gamma  \right)}} {\mbox{ .}}
\label{15}
\end{equation}
Assuming 
$h _{_{\left(e ^{\pm } \right)}}/h _{_{\left(\gamma  \right)}}$ to be given by their (constant) initial ratio, Eq. (\ref{10})  integrates to 
\begin{equation}
N _{_{\left(\gamma  \right)}}\left(t \right) 
= N _{_{\left(\gamma  \right)}}\left(t _{0}\right) 
+ \frac{h _{_{\left(e ^{\pm } \right)}}}{h _{_{\left(\gamma  \right)}}}
\left[N _{_{\left(e ^{\pm } \right)}}\left(t _{0} \right) 
- N _{_{\left(e ^{\pm } \right)}}\left(t  \right)\right]\ .
\label{16}
\end{equation}
The final value $N _{_{\left(\gamma \right)}}\left(t _{f} \right)$ corresponds to the case where the electrons and positrons have been annihilated, i.e., 
$N _{_{\left(e ^{\pm } \right)}}\left(t _{f} \right)=0$: 
\begin{equation}
\frac{N _{_{\left(\gamma  \right)}}\left(t _{f} \right)}
{N _{_{\left(\gamma  \right)}}\left(t _{0} \right)} 
= 1 + \frac{h _{_{\left(e ^{\pm } \right)}}}{h _{_{\left(\gamma  \right)}}}
\frac{n _{_{\left(e ^{\pm } \right)}}\left(t _{0} \right)}
{n _{_{\left(\gamma \right)}}\left(t _{0} \right)}\ .
\label{17}
\end{equation}
This result has to be coupled to the temperature law (\ref{9}), which for photons becomes 
\begin{equation}
\frac{\dot{T}_{_{\left(\gamma  \right)}}}{T _{_{\left(\gamma  \right)}}} 
= -\frac{\dot{a}}{a} 
+ \frac{1}{3}\frac{\dot{N}_{_{\left(\gamma  \right)}}}
{N _{_{\left(\gamma  \right)}}}
\quad\Rightarrow\quad
T _{_{\left(\gamma  \right)}}\left(t \right) 
= T _{_{\left(0\right)}}\frac{a \left(t _{0} \right)}{a \left(t \right)}
\left(\frac{N _{_{\left(\gamma  \right)}}\left(t \right)}
{N _{_{\left(\gamma  \right)}}\left(t _{0}\right)} \right)^{1/3}\ .
\label{18}
\end{equation}
For $t \geq t _{f}$ the ratio 
$N _{_{\left(\gamma  \right)}}\left(t \right)/
N _{_{\left(\gamma  \right)}}\left(t _{0}\right)$ is fixed by the value 
$N _{_{\left(\gamma  \right)}}\left(t _{f}\right)/
N _{_{\left(\gamma  \right)}}\left(t _{0}\right)$. 
Since  
$h _{_{\left(e ^{\pm } \right)}}/h _{_{\left(\gamma  \right)}}=7/6$ 
[cf. (\ref{15})]  
and 
$n _{_{\left(e ^{\pm } \right)}}\left(t _{0} \right)
/n _{_{\left(\gamma  \right)}}\left(t _{0}\right) = 3/2$ 
[cf. (\ref{14})] we obtain 
$N _{_{\left(\gamma  \right)}}\left(t _{f}\right)/
N _{_{\left(\gamma  \right)}}\left(t _{0} \right)=11/4$. 
Consequently, the temperature evolution law for $t \geq t _{f}$  is  
\begin{equation}
T _{_{\left(\gamma  \right)}}\left(t \right) 
= T _{_{\left(0\right)}}\frac{a \left(t _{0} \right)}{a \left(t \right)}
\left(\frac{11}{4} \right)^{1/3}
\quad\Rightarrow\quad 
\frac{T _{_{\left(\gamma  \right)}}\left(t \right)}
{T _{_{\left(\nu  \right)}}\left(t \right)} 
= \left(\frac{11}{4} \right)^{1/3}
\mbox{\ \ \ }\mbox{\ \ \ }\mbox{\ \ \ }
\left(t \geq t _{f} \right)\ .
\label{19}
\end{equation}
Thus we have reproduced the well-known difference between photon and neutrino background temperatures on the basis of the temperature law (\ref{9}).   
Usually, this result is obtained by calculating the entropy transfer from the 
$e ^{\pm }$ pairs to the photons under the condition of entropy conservation 
\cite{Boe}. 

\section{Evaporation of primordial black holes}
\label{PBH}

In a variety of scenarios with copious production of primordial black holes  the latter may substantially contribute to the energy density of the universe (see, e.g., \cite{ZPPBH}). 
Some of these models are characterized by a narrow mass spectrum \cite{BCL}. 
Under such circumstances it is a good approximation to ascribe the same mass 
$m _{_{\left(BH \right)}}$ to all members of the population. 
On the other hand, a black hole mass is known to be characterized by a temperature $T _{_{\left(BH \right)}}\propto m _{_{\left(BH \right)}}^{-1}$. 
Consequently, with a single mass population of PBHs one may associate a single temperature $T _{_{\left(BH \right)}}$   as well. 
Furthermore, one may show that a PBH population in the expanding universe may be regarded as an ensemble of non-interacting particles \cite{ZPPBH}. 
These properties suggest a description of the PBH component as a pressureless ``fluid'', in which $T _{_{\left(BH \right)}}$ in some respect plays the role of a fluid temperature. 
Since PBHs are known to evaporate, it is tempting to establish a two-fluid model along the lines of Sec. \ref{basic} with one component being the PBH ``fluid'', the second one radiation. 
The equations of state (\ref{7})  for the PBH component  are 
\begin{equation}
p _{_{\left(BH \right)}} = 0 \ ,
\mbox{\ \ }\mbox{\ \ }\mbox{\ \ }
\rho _{_{\left(BH \right)}} = n _{_{\left(BH \right)}}
m _{_{\left(BH \right)}}\ .
\label{20}
\end{equation}
The black hole temperature is related to its mass by
the well-known formula \cite{HAW2}
\begin{equation}
T _{_{\left(BH \right)}} = \frac{1}{8 \pi m _{_{\left(BH \right)}}}
\ .
\label{25}
\end{equation}  
This temperature is attributed to each  PBH individually, i.e., primarily it is {\it not} a conventional fluid  
temperature. 
The number $N _{_{\left(BH \right)}}$ of PBHs in a comoving volume  
$a ^{3}$,
$N _{_{\left(BH \right)}} = n _{_{\left(BH \right)}}a ^{3}$, is not  
preserved and, according to Eq. (\ref{2}), we may write down a  
balance equation for the corresponding PBH number flow vector
$N _{_{\left(BH \right)}}^{i} = n _{_{\left(BH \right)}}u ^{i}$,
\begin{equation}
N _{_{\left(BH \right)} ;i}^{i} =
\dot{n}_{_{\left(BH \right)}} + 3H n_{_{\left(BH \right)}} =
n _{_{\left(BH \right)}} \Gamma _{_{\left(BH \right)}}
{\mbox{ . }}
\label{21}
\end{equation}
The black hole energy balance becomes [cf. Eq. (\ref{2}) with (\ref{1})]
\begin{equation}
\dot{\rho }_{_{\left(BH \right)}} + 3H \rho _{_{\left(BH \right)}} = 
u _{a}t ^{a}_{_{\left(BH \right)}} = \rho _{_{\left(BH \right)}}
\left[\Gamma _{_{\left(BH \right)}}
+ \frac{\dot{m}_{_{\left(BH \right)}}}
{m _{_{\left(BH \right)}}}\right] \ .
\label{23}
\end{equation}
Using $p _{_{\left(BH \right)}}=0$ as well as Eq. (\ref{6}) in the fluid temperature law (\ref{8})  we find 
\begin{equation}
\dot{T}_{_{\left(BH \right)}}  =
\frac{u _{a} t _{_{\left(BH \right)}}^{a}
- \Gamma _{_{\left(BH \right)}} \rho _{_{\left(BH \right)}}}
{\partial \rho _{_{\left(BH \right)}}/ \partial T _{_{\left(BH  
\right)}}} 
= \frac{ \rho _{_{\left(BH \right)}}}
{\partial \rho _{_{\left(BH \right)}}/ \partial T _{_{\left(BH  
\right)}}}\frac{\dot{m}_{_{\left(BH \right)}}}
{m _{_{\left(BH \right)}}}
{\mbox{ .}}
\label{24a}
\end{equation}
We emphasize that we have used here the same symbol, 
$T _{_{\left(BH \right)}}$, for the the PBH ``fluid'' temperature and for the temperature (\ref{25}), which is ascribed to the individual black holes. 
The consistency of this identification becomes obvious if we combine the equations of state (\ref{20}) with (\ref{25}) and introduce the result for 
$\partial \rho _{_{\left(BH \right)}}/ \partial T _{_{\left(BH  
\right)}}$ into (\ref{24a}):
\begin{equation}
\frac{\partial{\rho _{_{\left(BH \right)}}}}
{\partial{T _{_{\left(BH \right)}}}} 
= - \frac{\rho _{_{\left(BH \right)}}}{T _{_{\left(BH \right)}}}
\quad\Rightarrow\quad
\dot{T}_{_{\left(BH \right)}}  
= - T _{_{\left(BH \right)}} \frac{\dot{m}_{_{\left(BH \right)}}}
{m _{_{\left(BH \right)}}}
{\mbox{ .}}
\label{24}
\end{equation}
It is the crucial point of our analysis that Hawking's temperature law (\ref{25}) for individual black holes fits together with the general fluid temperature law (\ref{8}) for the equations of state (\ref{20}) with (\ref{25}).  
This circumstance provides the basis for our thermodynamical discussion of the PBH evaporation process. 
To this purpose we identify component 1 of the general analysis in section \ref{basic}  with the PBH ``fluid'' and component 2 with ulrarelativistic matter (radiation, subscript r), i.e., $\left(1 \right)\rightarrow \left(BH \right)$ and 
$\left(2 \right)\rightarrow \left(r \right)$. For the latter we require constant entropy per particle, i.e., [cf. Eq. (\ref{6})]
\begin{equation}
\dot{s}_{_{\left(r \right)}}= 0 
\quad\Rightarrow\quad 
u _{a}t ^{a}_{_{\left(r \right)}} 
= \frac{4}{3}\rho _{_{\left(r \right)}}\Gamma _{_{\left(r \right)}} 
\quad\Rightarrow\quad
\frac{\dot{T}_{_{\left(r \right)}}}{T _{_{\left(r \right)}}} 
= - H \left(1 - \frac{\Gamma _{_{\left(r \right)}}}{3H}\right)\ .
\label{26}
\end{equation}
Combination with 
$t ^{a}_{_{\left(BH \right)}}=-t ^{a}_{_{\left(r \right)}}$ 
from (\ref{4}) yields 
\begin{equation}
\Gamma _{_{\left(r \right)}} = - \frac{4}{3}
\frac{\rho _{_{\left(BH \right)}}}{\rho _{_{\left(r \right)}}}
\left[\Gamma _{_{\left(BH \right)}}
+ \frac{\dot{m}_{_{\left(BH \right)}}}
{m _{_{\left(BH \right)}}}\right]\ .
\label{27}
\end{equation}
The total entropy flow $S ^{a}$ is the sum of the contributions 
$S ^{a}_{_{\left(BH \right)}}
=n_{_{\left(BH \right)}}s _{_{\left(BH \right)}}u ^{a}$ and 
$S ^{a}_{_{\left(r\right)}}
=n_{_{\left(r \right)}}s _{_{\left(r\right)}}u ^{a}$. 
With $s _{_{\left(BH \right)}} = 4 \pi  m _{_{\left(BH \right)}}^{2}$  we obtain the following expression for the overall entropy production density \cite{ZPPBH}: 
\begin{equation}
S ^{a}_{; a} = \rho _{_{\left(BH \right)}} \Gamma _{_{\left(BH  
\right)}}
\left[\frac{1}{2 T _{_{\left(BH \right)}}}
- \frac{1}{T _{_{\left(r \right)}}}\right]
+ \rho _{_{\left(BH \right)}} \frac{\dot{m}_{_{\left(BH \right)}}}
{m _{_{\left(BH \right)}}}
\left[\frac{1}{T _{_{\left(BH \right)}}}
- \frac{1}{T _{_{\left(r \right)}}}\right]\ .
\label{28}
\end{equation}
It is obvious from (\ref{27})  that negative values of 
$\dot{m}_{_{\left(BH \right)}}$ (and $\Gamma _{_{\left(BH \right)}}$) correspond to a positive quantity $\Gamma _{_{\left(r \right)}}$. 
This case is equivalent to the creation of radiative particles at the expense of the PBH mass (and its number), i.e., to PBH evaporation. 
The inverse process, namely ``accretion'' with  
$\dot{m}_{_{\left(BH \right)}}>0$ and $\Gamma _{_{\left(r \right)}}<0$ is described by the general formula (\ref{28}) as well. 
Which of the two processes is thermodynamically preferred depends on the ratio of the temperatures. 
Given a specific initial ratio, the further evolution is entirely governed by the temperature laws in (\ref{24})  and (\ref{26}). 
Let's assume an initial configuration with 
$T_{_{\left(BH \right)}}\left(t _{0} \right) 
=T_{_{\left(r \right)}}\left(t _{0} \right)$. A non-negative entropy production density then requires 
$\Gamma _{_{\left(BH \right)}}\leq 0$. 
Since one expects 
$\dot{m}_{_{\left(BH \right)}}$ and $\Gamma _{_{\left(BH \right)}}$ to have the same sign, this implies a positive value of 
$\Gamma _{_{\left(r \right)}}$, i.e., radiation particles are produced which makes the PBH masses shrink. 
The further evolution depends on a subtle interplay between the rates 
$\Gamma _{_{\left(BH \right)}}$ and $\Gamma _{_{\left(r \right)}}$ and their respective back reactions on the temperature laws (\ref{24})  and (\ref{26}). 
A positive $\Gamma _{_{\left(r \right)}}$ may either be  
larger or smaller than the expansion rate $3H $. 
For $\Gamma _{_{\left(r\right)}} < 3H $ the fluid temperature  
decreases according to Eq. (\ref{26}), while the BH temperature  
increases according to Eq. (\ref{24}).
It follows that $T _{_{\left(BH \right)}} > T _{_{\left(r \right)}}$ 
at $t > t _{0}$. The evaporation process will continue since
$T _{_{\left(r \right)}} < T _{_{\left(BH \right)}}$ requires
$\Gamma _{_{\left(BH \right)}} < 0$ and
$\dot{m}_{_{\left(BH \right)}}/m _{_{\left(BH \right)}} < 0$ to  
guarantee
$S ^{a}_{;a} > 0$ in Eq. (\ref{28}). 
For $\Gamma _{_{\left(r \right)}} >
3H $, however, hypothetically realized e.g. by a large initial ratio 
$\rho _{_{\left(BH \right)}}/ \rho _{_{\left(r \right)}}$,
the fluid temperature increases.
If this increase is smaller than the increase in $T _{_{\left(BH  
\right)}}$
we have again
$T _{_{\left(r\right)}} < T _{_{\left(BH \right)}}$ and the PBH  
evaporation goes on since it remains thermodynamically favored 
($S ^{a}_{;a} > 0$).
But an increase in $T _{_{\left(r \right)}}$ stronger than that in
$T _{_{\left(BH \right)}}$ results in a fluid temperature which is  
higher than $T _{_{\left(BH \right)}}$.
For $T _{_{\left(r \right)}} > 2 T _{_{\left(BH \right)}}$ a  
positive entropy production (\ref{28}) requires
$\Gamma _{_{\left(BH \right)}} > 0$ and
$\dot{m}_{_{\left(BH \right)}}/m _{_{\left(BH \right)}} > 0$,   
implying a quick transition to a negative 
$\Gamma _{_{\left(r\right)}}$, i.e., the process can no longer continue.
A strong ``reheating'' of the fluid will stop the evaporation and  
reverse the process.
Now, the second law requires PBHs to be formed out of the radiation  
and to accrete mass.
A negative $\Gamma _{_{\left(r \right)}}$, on the other hand, will make 
$T _{_{\left(r \right)}}$ subsequently decrease [cf. Eq. (\ref{26})]. 
If $T _{_{\left(r \right)}}$ has fallen below 
$T _{_{\left(BH \right)}}$, the evaporation process may set in again.  
In particular, this self-confining property implies that a catastrophic growth of the PBHs is thermodynamically forbidden. 
The point is that 
a PBH growth, i.e. mass accretion with  
$\dot{m}_{_{\left(BH \right)}}/m _{_{\left(BH \right)}} > 0$, 
back reacts on the temperature of the ambient radiation.  
For a fixed PBH number, i.e. $\Gamma _{_{\left(BH \right)}}=0$, the corresponding radiation temperature changes as 
\[
\frac{\dot{T}_{_{\left(r \right)}}}{T _{_{\left(r \right)}}} = - H
\left(1 + \frac{1}{4}\frac{\rho _{_{\left(BH \right)}}}
{\rho _{_{\left(r \right)}}}
\frac{\dot{m}_{_{\left(BH \right)}}}{m _{_{\left(BH \right)}}}
H ^{-1}\right) \ .
\]
It is obvious that for $\dot{m}_{_{\left(BH \right)}}>0$ from some time on the temperature 
$T _{_{\left(r \right)}}$ will
cool off faster than $T _{_{\left(BH \right)}}$ [cf. Eq. (\ref{24})]. 
Consequently, $T _{_{\left(r \right)}}$ will approach 
$T _{_{\left(BH \right)}}$. 
As soon as $T _{_{\left(r \right)}}$ has fallen below 
$T _{_{\left(BH \right)}}$, mass accretion stops
since for $T _{_{\left(r\right)}} < T _{_{\left(BH \right)}}$ the rate 
$\dot{m}_{_{\left(BH \right)}}/m _{_{\left(BH \right)}}$ has to be negative in order to guarantee a
positive entropy production, i.e., the process now proceeds in the
reverse direction and the PBHs can no longer grow but start to evaporate again.  
This completes our thermodynamic discussion of PBH evaporation based on the temperature law (\ref{8}) (and its special case (\ref{9})).  

\section{``Deflationary'' universe}
\label{defl}

In the two previous examples we did not consider the impact of the decay and production processes on the expansion behavior of the universe. 
As was shown in \cite{ZPepa} and \cite{ZPPBH}, the general tendency of this influence is to increase the cosmic expansion rate. Namely, processes of the type discussed in sections \ref{epa} and \ref{PBH} give rise to an effective viscous pressure of the cosmic medium as a whole [cf. Eq. (\ref{4})]. Since this contribution to the overall pressure  
is negative, its net effect is to accelerate the expansion. 
While this effect is small for the cases dealt with in sections \ref{epa} and \ref{PBH}, it is essential in the ``deflationary'' universe model of the present section.  
This model    
relies on Einstein's field equations with the energy-momentum tensor (\ref{4})  of a bulk viscous fluid. In a homogeneous and isotropic universe one  has  
\begin{equation}
\kappa \rho = 3 H ^{2}\ ,
\mbox{\ \ \ \ }
\dot{H} = - \frac{\kappa}{2}\left(\rho + p + \Pi  \right)
\quad\Rightarrow\quad
\kappa \Pi = - 3 \gamma H ^{2} - 2 \dot{H}\ ,
\label{29}
\end{equation} 
where $\kappa$ is Einstein's gravitational constant and 
$\gamma \equiv  1+p/ \rho $. 
In case $\Pi $ is not a ``conventional'' viscous pressure but represents a quantity describing cosmological particle production on a phenomenological level \cite{Prig,Calv}, it may be related to the production rate $\Gamma $  introduced in (\ref{5}). 
For ``adiabatic'' particle production this relation is 
\begin{equation}
\Pi = - \left(\rho + p \right)\frac{\Gamma }{3H}\ .
\label{30}
\end{equation}
Combination with the field equations (\ref{29}) then yields \cite{Zpreprint}
\begin{equation}
\frac{\Gamma }{3H} = 1 + \frac{2}{3 \gamma }
\frac{\dot{H}}{H ^{2}} 
\quad\Rightarrow\quad
\frac{H ^{\prime }}{H \left[\frac{\Gamma }{3H} - 1 \right]} 
= \frac{3}{2}\frac{\gamma }{a}
\ ,
\label{31} 
\end{equation}
where $H ^{\prime } \equiv  \mbox{d}H/ \mbox{d}a$. 
Strictly speaking, the rate $\Gamma $ has to be calculated on the quantum level (see, e.g., \cite{SchDe1,SchDe2}).  
In a phenomenological setting an ansatz for $\Gamma /H$ is required.  
For a dependence $\Gamma \propto \rho \propto H ^{2}$ 
\cite{GunzMaNe,LiMa,Zpreprint}
and $\gamma =4/3$ we obtain 
\begin{equation}
H = 2\frac{a _{e} ^{2}}
{a ^{2} + a _{e} ^{2}}H _{e}
\quad\Rightarrow\quad
\frac{\Gamma }{3H} 
= \frac{a _{e} ^{2}}
{a ^{2} + a _{e} ^{2}}
\ ,
\label{32}
\end{equation}
where we have chosen the constants such that 
$\dot{H}_{e} = - H _{e}^{2}$, i.e, $\ddot{a}>0$ for $a<a _{e}$ and 
$\ddot{a}<0$ for $a>a _{e}$. 
$H$ starts with a constant value $H _{0}=2H _{e}$ at $a \ll a _{e}$ and then ``deflates''  towards the typical $H \propto a ^{-2}$ behaviour of a radiation dominated universe for $a \gg a _{e}$.  
This Hubble rate has originally been obtained in the context of phenomenological approaches to cosmological vacuum decay \cite{GunzMaNe,LiMa}. 
Again, this is a two-component model with one component playing the role of the cosmological ``vacuum''. 
Our point here is to demonstrate that such kind of model fits into the general structure of section \ref{basic} and admits a similar thermodynamic discussion  as the cases of electron-positron annihilation and PBH evaporation. 
We will identify 
the first component of the general formalism in Sec. \ref{basic} with the ``vacuum'' (subscript $v$), i.e., 
$\left(1 \right)\rightarrow \left(v \right)$,  the second one again with radiation, i.e., $\left(2 \right)\rightarrow \left(r \right)$. The  sketched 
scenario may then be obtained on the basis of an interacting two-fluid model with $\rho = \rho _{_{\left(v \right)}} + \rho _{_{\left(r \right)}}$ where  
\begin{equation}
\rho _{_{\left(v \right)}} = \frac{3 H _{e}^{2}}{2 \pi }m _{P}^{2}
\left[\frac{a _{e}^{2}}{a ^{2} + a _{e}^{2}} \right]^{3}\ ,
\mbox{\ }\mbox{\ }\mbox{\ }
\rho _{_{\left(r \right)}} = \frac{3 H _{e}^{2}}{2 \pi }m _{P}^{2}
\left(\frac{a}{a _{e}} \right)^{2}
\left[\frac{a _{e}^{2}}{a ^{2} + a _{e}^{2}} \right]^{3}\ ,
\label{33}
\end{equation}
and $m _{P}^{2}=8 \pi / \kappa$ is the square of the Planck mass.   
The part $\rho _{_{\left(v \right)}}$ is finite for $a \rightarrow 0$ and decays as $a ^{-6}$ for $a \gg a _{e}$, while 
the part $\rho _{_{\left(r \right)}}$ describes relativistic matter with 
$\rho _{_{\left(r \right)}}\rightarrow 0$ for $a \rightarrow 0$ and 
$\rho _{_{\left(r \right)}}\propto a ^{-4}$  for $a \gg a _{e}$. 
The energy balances are ($A = v, r$)
\begin{equation}
\dot{\rho }_{_{\left(A \right)}} + 3H \left[\rho _{_{\left(A \right)}} 
+ p _{_{\left(A \right)}}\right] 
= \Gamma _{_{\left(A \right)}}
\left[\rho _{_{\left(A \right)}} + p _{_{\left(A \right)}}\right]
\label{34}
\end{equation}
with
\begin{equation}
\frac{\Gamma _{_{\left(v \right)}}}{3H} 
= \left(1 - \frac{1}{2}\frac{a ^{2}}{a _{e}^{2}} \right)
\frac{a _{e}^{2}}{a ^{2} + a _{e}^{2}}\ ,
\mbox{\ \ \ }\mbox{\ \ \ }\mbox{\ \ \ }
\frac{\Gamma _{_{\left(r \right)}}}{3H} = \frac{3}{2}\frac{a _{e}^{2}}{a ^{2} + a _{e}^{2}}\ .
\label{35}
\end{equation}
The equation for $\rho _{_{\left(v \right)}}$ may be written as 
\begin{equation}
\dot{\rho} _{_{\left(v \right)}} 
+ 3H \left(\rho _{_{\left(v \right)}} 
+ p _{_{\left(v \right)}} + \Pi _{_{\left(v \right)}}\right) = 0 \ ,
\label{36}
\end{equation}
where
\begin{equation}
\Pi _{_{\left(v \right)}} \equiv  
- \frac{\Gamma _{_{\left(v \right)}}}
{3H}\left(\rho _{_{\left(v \right)}} + p _{_{\left(v \right)}} \right) 
= - \left(1 - \frac{1}{2}\frac{a ^{2}}{a _{e}^{2}} \right)
\frac{a _{e}^{2}}{a ^{2} + a _{e}^{2}}
\left(\rho _{_{\left(v \right)}} + p _{_{\left(v \right)}} \right)\ .
\label{37}
\end{equation}
This corresponds to an effective equation of state 
\begin{equation}
P _{_{\left(v \right)}} \equiv  p _{_{\left(v \right)}} 
+ \Pi _{_{\left(v \right)}} 
= \frac{a ^{2} - a _{e}^{2}}{a ^{2} + a _{e}^{2}}\rho _{_{\left(v \right)}}\ .
\label{38}
\end{equation}
Although we have always 
$p _{_{\left(v \right)}}= \rho _{_{\left(v \right)}}/3$, 
the effective equation of state for $a \rightarrow 0$ approaches 
$P _{_{\left(v \right)}} = - \rho _{_{\left(v \right)}}$. 
Effectively, this component behaves as a vacuum contribution. 
For $a \gg a _{e}$  it represents stiff matter with 
$P _{_{\left(v \right)}} = \rho _{_{\left(v \right)}}$.  
The radiation component may be regarded as emerging from the decay of the initial vacuum according to 
\begin{equation}
\dot{\rho }_{_{\left(r \right)}} + 4H \rho _{_{\left(r \right)}} 
= - \dot{\rho }_{_{\left(v \right)}}\ .
\label{39}
\end{equation}
The radiation temperature is obtained from the general law (\ref{9}), which in the present case specifies to 
\begin{equation}
\frac{\dot{T}_{_{\left(r \right)}}}{T _{_{\left(r \right)}}} 
= - H \left(1 - \frac{\Gamma _{_{\left(r \right)}}}{3H} \right) 
= - H \left[1 - \frac{3}{2}\frac{a _{e}^{2}}{a ^{2} + a _{e}^{2}} \right]\ .
\label{40}
\end{equation}
Integration yields 
\begin{equation}
T _{_{\left(r \right)}} = 2 ^{3/4}T _{_{\left(e,r \right)}}\frac{a _{e}}{a}
\left[\frac{a ^{2}}{a ^{2} + a _{e}^{2}} \right]^{3/4}
\quad\Rightarrow\quad
T _{_{\left(r \right)}}\propto \rho _{_{\left(r \right)}}^{1/4}\ .
\label{41}
\end{equation}
$T _{_{\left(e,r \right)}}$ is the value of the radiation temperature at 
$a=a _{e}$.  
This temperature starts at $T _{_{\left(r \right)}}=0$ for $a=0$, then increases to  a maximum value, given by $\Gamma _{_{\left(r \right)}} = 3H$, equivalent to 
$a ^{2}= \frac{1}{2}a _{e}^{2}$, 
\begin{equation}
T _{_{\left(r \right)}}^{max} = \left(\frac{32}{27} \right)^{1/4}
T _{_{\left(e,r \right)}}\ , 
\label{42}
\end{equation}
and finally decreases as $a ^{-1}$ for large values of $a$.  
Our formalism allows us to ascribe a temperature $T _{_{\left(v \right)}}$ to the ``vacuum'' as well, which is determined analogously by 
\begin{equation}
\frac{\dot{T}_{_{\left(v \right)}}}{T _{_{\left(v \right)}}} 
= - H \left(1 - \frac{\Gamma _{_{\left(v \right)}}}{3H} \right) 
= - \frac{3}{2}H \frac{a ^{2}}{a ^{2}+a _{e}^{2}}\ .
\label{43}
\end{equation}
The ``vacuum'' temperature  behaves as
\begin{equation}
T _{_{\left(v \right)}} 
= T _{_{0}}\left[\frac{a _{e}^{2}}{a ^{2}+a _{e}^{2}} \right]^{3/4}\ .
\label{44}
\end{equation}
It starts from a maximum value at $a=0$ and decreases as $a ^{-3/2}$ for large $a$. 
The ``vacuum'' is radiative in the sense that 
$\rho _{_{\left(v \right)}}\propto T _{_{\left(v \right)}}^{4}$ is valid. 
As a final remark we mention that it is also possible to introduce a temperature $T$ of the cosmic medium as a whole with a behavior \cite{Zpreprint}
\begin{equation}
\frac{\dot{T}}{T} 
= - H \left(1 - \frac{\Gamma }{3H} \right) 
= - H \frac{a ^{2}}{a ^{2}+a _{e}^{2}}
\quad\Rightarrow\quad
T = T _{_{0}}
\left[\frac{a _{e}^{2}}{a ^{2} + a _{e}^{2}} \right]^{1/2} 
 \ ,
\label{45}
\end{equation}
which ``interpolates'' between (\ref{44}) for small $a$ and (\ref{41})  for 
$a \gg a _{e}$.  
These considerations clarify the central role played by the general temperature law (\ref{9}) also under conditions where the relevant back reaction substantially affects the entire cosmological dynamics. 

\section{Conclusions}
\label{conclusions}

Cosmological thermodynamics allows us to establish a unifying view on a broad range of different phenomena and to uncover joint underlying structures. 
In this paper we have explored similar thermodynamic features of matter creation in the early universe, primordial black hole evaporation, and electron-positron annihilation after neutrino decoupling.  
All these processes are governed by the same basic temperature law for a fluid with variable particle number which 
takes into account the back reaction of the 
relevant interactions on the thermal history of the universe. 
A particular aspect of our considerations is the consistency of this law     
with Hawking's black hole temperature formula  
$T _{_{\left(BH \right)}} \propto m _{_{\left(BH \right)}}^{-1}$.  
This circumstance provides the basis for a two-fluid model for the evaporation of a single-mass PBH component into radiation. 

\ \\
{\bf Acknowledgment}\\
This paper was supported by the
Deutsche Forschungsgemeinschaft,
the Spanish Ministry of Education
(grant PB94-0718) and NATO
(grant CRG 940598).

\small
\ \\

\ \\

\end{document}